\documentclass[twocolumn,showpacs,prd]{revtex4}
\usepackage{mathrsfs,bm}
\usepackage{longtable,lscape}
\usepackage{txfonts}
\usepackage{amssymb}
\usepackage{indentfirst}
\usepackage{graphicx,,booktabs}
\usepackage{color}
\usepackage{amssymb}

\begin{document}

\title{Non-strange partner of strangeonium-like state $Y(2175)$}
\author{Xiao Wang$^{1,2}$}
\author{Zhi-Feng Sun$^{1,2}$}
\author{Dian-Yong Chen$^{1,3}$}
\author{Xiang Liu$^{1,2}$\footnote{Corresponding author}}\email{xiangliu@lzu.edu.cn}
\author{Takayuki Matsuki$^4$}\email{matsuki@tokyo-kasei.ac.jp}

\affiliation{
$^1$Research Center for Hadron and CSR Physics,
Lanzhou University and Institute of Modern Physics of CAS, Lanzhou 730000, China\\
$^2$School of Physical Science and Technology, Lanzhou University, Lanzhou 730000,  China\\
$^3$Nuclear Theory Group, Institute of Modern Physics of CAS, Lanzhou 730000, China\\
$^4$Tokyo Kasei University, 1-18-1 Kaga, Itabashi, Tokyo 173-8602, Japan}
\date{\today}

\begin{abstract}
Inspired by the observed $Y(2175)$ state, we predict its
non-strange partner $Y(1915)$, which has a resonance structure
with mass around 1915 MeV and width about $317\sim 354$ MeV.
Experimental search for $Y(1915)$ is proposed by analyzing the
$\omega f_0(980)$ or $\omega \pi\pi$ invariant mass spectrum of
the $e^+e^-\to \omega f_0(980), \omega \pi\pi$ and $J/\psi\to \eta
\omega f_0(980)$ processes, which are accessible at Belle, BaBar,
BESIII and forthcoming BelleII. Considering similarity between two
families, the comparison of the mass spectra of $\omega$ and
$\phi$ families can provide important information on the $1D$
state of $\phi$ family, $\phi(1910)$, which has a very broad
resonance structure with mass around 1910 MeV regarded as the
strangeonium partner of $\omega(1650)$. This also answers the
question why the $1D$ state $\phi(1910)$ is still missing in
experiment. This is supported by our former study on the
properties of $Y(2175)$, which explains $Y(2175)$ as the $2D$
strangeonium because our theoretical total width is comparable with
the Belle data.

\end{abstract}

 \maketitle
%\end{CJK}

The BaBar Collaboration first reported the observation of strangeonium-like state $Y(2175)$ in $e^+e^-\to \phi(1020) f_0(980)$ \cite{Aubert:2006bu}, which was later confirmed by BESII in $J/\psi\to\eta\phi(1020) f_0(980)$ \cite{:2007yt} and by Belle in the
$e^+e^-\to \phi(1020) f_0(980),\phi(1020)\pi^+\pi^-$ processes \cite{Shen:2009zze}. As the only one $XYZ$ state with light flavor, $Y(2175)$ has stimulated theorists to consider various theoretical explanations (see Refs. \cite{Ding:2007pc,Coito:2009na,Ding:2006ya,MartinezTorres:2008gy,Wang:2006ri,Chen:2008ej,Klempt:2007cp,Qiao:2005av} formore details).

What is more important for the observation of $Y(2175)$ is that $Y(2175)$, combined with the observed $Y(4260)$ \cite{Aubert:2005rm} and $\Upsilon(10860)$ \cite{Abe:2007tk}, forms a complete series of flavors \cite{Chen:2011cj}, which is mainly due to the similarity among decay processes of these particles,
%existing in these observed phenomena related to $Y(2175)$, $Y(4260)$ and $\Upsilon(10860)$
\begin{eqnarray}
e^+e^-\rightarrow
\left\{
  \begin{array}{cccc}
    Y(2175) & \rightarrow  & \phi(1020)\pi^+\pi^-&\mathrm{strange}\\
    Y(4260) & \rightarrow  & J/\psi\pi^+\pi^-&\mathrm{charm}\\
    \Upsilon(10860)& \rightarrow  & \Upsilon(1S,2S)\pi^+\pi^-&\mathrm{bottom}\\
  \end{array}
\right. .\label{common}
\end{eqnarray}
A natural deduction from Eq. (\ref{common}) is that there must exist a non-strange counterpart of $Y(2175)$, which can decay into $\omega(782)\pi^+\pi^-$.
However, until now such a non-strange partner of $Y(2175)$ is still missing in experiment. Thus, the study of the non-strange partner of $Y(2175)$ can not only provide valuable information on further experimental search, but will also be helpful to reveal the underlying physics of $Y(2175)$. It is no doubt that this is an intriguing and important research topic.

\begin{figure}[htb]
\centering
\begin{tabular}{cc}
\scalebox{0.35}{\includegraphics{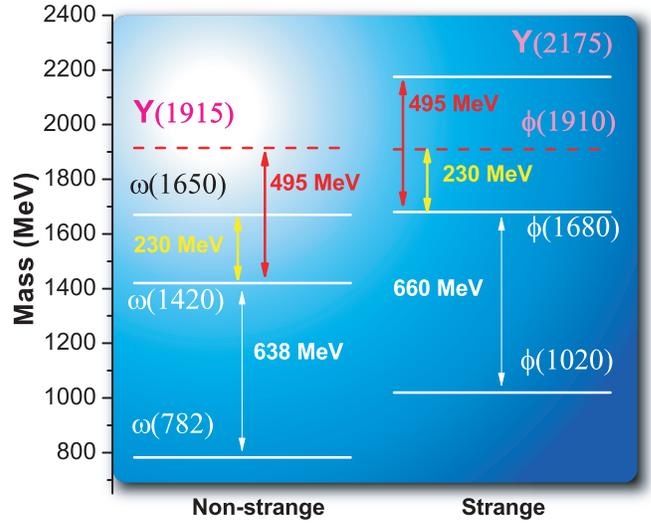}}
\end{tabular}
\caption{(color online). Comparison of isoscalar light vector mesons. Besides these well established $\omega(782)$, $\omega(1420)$, $\omega(1650)$, $\phi(1020)$, and $\phi(1680)$, the newly observed strangeonium-like state $Y(2175)$ with two predicted states $Y(1915)$ and $\phi(1910)$ are also listed here. }
\label{sp}
\end{figure}

{\it Spectroscopy}: If comparing the mass spectra of $\omega$ and $\phi$ meson families just shown in Fig. \ref{sp}, one notices that the mass gap of $\omega(782)$
and $\omega(1420)$ is similar to that of $\phi(1020)$ and $\phi(1680)$, where $\omega(782)/\phi(1020)$ and $\omega(1420)/\phi(1680)$ are $1S$ and
$2S$ states, respectively. In addition, the mass difference between $\omega(782)$ and $\phi(1020)$ ($\sim 240$ MeV) is only 20 MeV smaller than that between $\omega(1420)$ and $\phi(1680)$ ($\sim 260$ MeV). These phenomena reflect the similar dynamics describing $\omega$ and $\phi$ meson families. Thus, the study of $\omega$ and $\phi$ meson families can be borrowed from each other, which enables us to estimate masses of the missing states in these meson families.

Among these observed states in the $\omega$ meson family, $\omega(1650)$ is a good candidate for the $1D$ state \cite{Barnes:1996ff}, while the corresponding partner of $\omega(1650)$ is missing in the $\phi$ family. The mass gap between $\omega(1650)$ and $\omega(1420)$ ($\sim 230$ MeV) can be applied to estimate the mass of $1D$ state in $\phi$ family, which is around 1910 MeV, corresponding to our predicted $\phi(1910)$ in Fig. \ref{sp}.
If checking the Particle Data Group (PDG) data \cite{Nakamura:2010zzi}, we can find several observed vector mesons $\rho(1700)$, $K^*(1680)$, and $\omega(1650)$, which form a flavor nonet with our predicted $\phi(1910)$.

Adopting the same discussion as above, we can naturally estimate the mass of the non-strange partner of $Y(2175)$. If the mass difference between the non-strange partner of $Y(2175)$ and $\omega(1420)$ is the same as that between $Y(2175)$ and $\phi(1680)$ ($\sim 495$ MeV), the mass of the non-strange partner of $Y(2175)$ should be close to $1915$ MeV, which corresponds to $Y(1915)$ listed in Fig. \ref{sp}. The above estimate is based on the conventional quarkonium explanation for $Y(2175)$.

\begin{figure}[htb]
\centering
\begin{tabular}{cc}
\scalebox{0.5}{\includegraphics{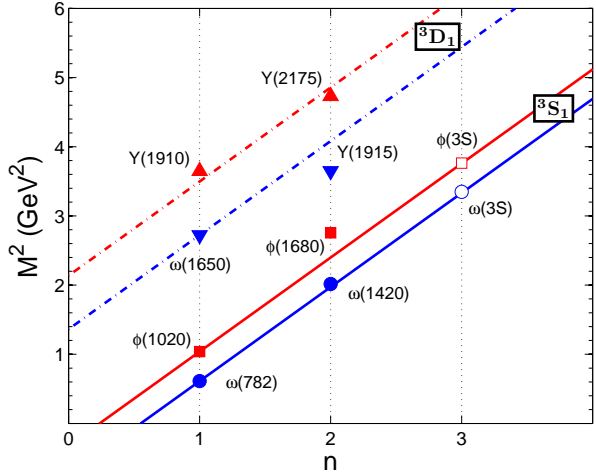}}
\end{tabular}
\caption{(color online). The analysis of the Regge trajectories.}
\label{rt}
\end{figure}

In addition, the quantitative analysis of the Regge trajectories seems to support these mass assignments, which is derived from the $I^G(J^{PC})=0^-(1^{--})$ trajectory on the $(n,M^2)$ plane with the relation $M^2=M_0^2+(n-1)\mu^2$ \cite{Chew:1962eu,Anisovich:2000kxa}, where $M_0$, $n$, and $\mu^2$ are the ground state mass, the radial quantum number, and the slope parameter of the trajectory, respectively (see Fig. \ref{rt}). The analysis of the Regge trajectories further indicates that $Y(2175)$ should be a $2D$ strangeonium while $3S$ assignment can be excluded. Thus, its non-strange partner $Y(1915)$ is a $2D$ state in the $\omega$ meson family.
The above analysis requires $\mu^2=1.36$ GeV$^2$ consistent with the range given in Ref. \cite{Anisovich:2000kxa}.

%Exclusion of $3S$ assignment to $Y(2175)$ is supported by Ref. \cite{Barnes:2002mu} since the predicted width, around $380$ MeV, for $3S$ vector strangeonium is far away from the measured width of $Y(2175)$ \cite{Aubert:2006bu,:2007yt,Shen:2009zze}.

{\it Decay}: As the non-strange partner of $Y(2175)$, $Y(1915)$
mainly decays into meson pairs. Thus, studying the strong decay of
$Y(1915)$ can provide valuable information on its total width and
partial decay widths, which will be helpful in further
experimental search for $Y(1915)$. Of course, we can easily extend
the same framework to calculate the strong decay of its strange
partner $Y(2175)$, which can test the $2D$ strangeonium assignment
to $Y(2175)$. Besides estimating the mass of $\phi(1910)$, in this
letter we also calculate the strong decay of $\phi(1910)$. This
study will answer the question why $\phi(1910)$ is still missing
in present experiment and how to search for it in future
experiment. In addition, we also present the results of
$\phi(1680)$, $\omega(1420)$ and $\omega(1650)$ decays, which can
test the reliability of the phenomenological model of strong
decay.

We adopt the quark pair creation model \cite{Micu:1968mk,LeYaouanc:1977gm,LeYaouanc:1988fx} to calculate the Okubo-Zweig-Iizuka-allowed strong decays of the states discussed above. For process ${A}(q(1)\bar{q}(2))\to {B}(q(1)\bar{q}(3))+{C}(q(4)\bar{q}(2))$, the transition matrix element reads as
$\langle BC|\mathcal{T}_{QPC}|A \rangle=\delta^3(\bm{K}_B+\bm{K}_C)\mathcal{M}^{M_{J_A}M_{J_B}M_{J_C}}(\bm{K})$. By the transition operator depicting the quark pair created from the vacuum
\begin{eqnarray}
\mathcal{T}_{QPC}&=& - 3 \gamma \sum_m\: \langle 1\;m;1\;-m|0\;0 \rangle\,
\int\!{\rm d}{\bm{k}}_3\; {\rm
d}{\bm{k}}_4 \delta^3({\bm{k}}_3+{\bm{k}}_4) \nonumber\\&&\times{\cal
Y}_{1m}\left(\frac{{\bm{k}}_3-{\bm{k}_4}}{2}\right)\;
\chi^{3 4}_{1, -\!m}\; \varphi^{3 4}_0\;\,
\omega^{3 4}_0\; d^\dagger_{3i}({\bm{k}}_3)\;
b^\dagger_{4j}({\bm{k}}_4)\,,\nonumber
\end{eqnarray}
we can deduce the concrete expressions of $\mathcal{M}^{M_{J_A}M_{J_B}M_{J_C}}(\bm{K})$. In the definition of $\mathcal{T}_{QPC}$, subscripts $i$ and $j$ are the $SU(3)$ color indices of the created quark
and anti-quark from the vacuum. $\chi_{{1,-m}}^{34}$ denotes a triplet
state of spin. $\mathcal{Y}_{\ell m}(\bm{k})\equiv
|\bm{k}|^{\ell}Y_{\ell m}(\theta_{k},\phi_{k})$ is the
$\ell$th solid harmonic polynomial. $\varphi^{34}_{0}=(u\bar u +d\bar d +s \bar
s)/\sqrt 3$ and $\omega_{0}^{34}=\delta_{\alpha_3\alpha_4}/\sqrt 3\,(\alpha=1,2,3)$ means flavor and
color singlets, respectively. Dimensionless
constant $\gamma=8.68$ is the strength of the quark pair creation from the
vacuum, which is determined by fitting the experimental data. The strength of $s\bar{s}$ creation satisfies
$\gamma_{s}=\gamma/\sqrt{3}$ \cite{LeYaouanc:1977gm}. By the relation \cite{Jacob:1959at}
\begin{eqnarray}
{\mathcal{M}}^{J L}(A\rightarrow BC) &=&
\frac{\sqrt{2 L+1}}{2 J_A +1} \!\! \sum_{M_{J_B},M_{J_C}} \langle
L 0 J M_{J_A}|J_A  M_{J_A}\rangle \nonumber\\&&\times\langle
J_B M_{J_B} J_C  M_{J_C} | J M_{J_A} \rangle \mathcal{M}^{M_{J_A}
M_{J_B} M_{J_C}}({\bm{K}}),\nonumber\label{JB}
\end{eqnarray}
we obtain the partial wave amplitude ${\mathcal{M}}^{J L}(A\rightarrow BC)$, where
$\bm{J}=\bm{J}_B+\bm{J}_C$ and
$\bm{J}_{A}+\bm{J}_{P}=\bm{J}_{B}+\bm{J}_C+\bm{L}$. Thus, the decay width is expressed as
$\Gamma=\pi^2|\bm{K}|\sum_{J,L}|{\mathcal{M}}^{J L}|^2/M_A^2$, where $|\bm{K}|$ is the three momentum of meson $B$ or $C$ in the center of mass frame of meson $A$. $M_A$ denotes the mass of meson $A$.
In calculating the spatial integral of the decay amplitude, the harmonic oscillator (HO) wave function $\Psi_{n_r\ell m}(\mathbf{k}) = \mathcal{R}_{n_r\ell}(R,\mathbf{k})\mathcal{Y}_{n_r\ell m}(\mathbf{k})$ is adopted to describe the meson wave function involved in the decays, where parameter $R$ in the HO wave function is obtained by reproducing the realistic root mean square radius by solving the Schr\"{o}dinger equation with the effective potential \cite{Close:2005se}.

\begin{center}
\begin{figure}[htbp!]
\includegraphics[bb=5 160 700 680,scale=0.55,clip]{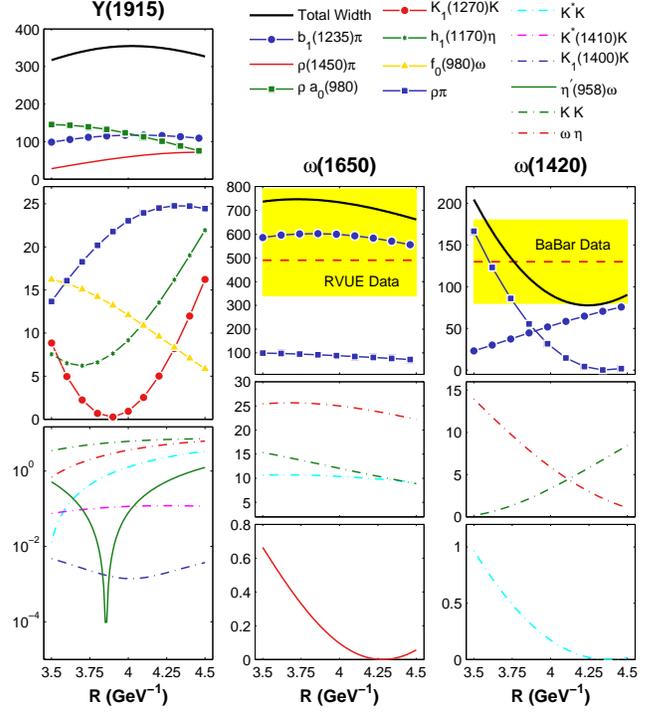}
\caption{(color online). The total (black solid line) and partial decay widths of $Y(1915)$, $\omega(1650)$ and $\omega(1420)$
and the comparison with the experimental data (red dashed line with yellow band).
\label{Fig: width1}}
\end{figure}
\end{center}

In Fig. \ref{Fig: width1}, we show the total and partial decay
widths of the predicted $Y(1915)$ dependent on $R$. Its total
decay width can reach up to $317\sim354$ MeV corresponding to the
range $R=3.5\sim 4.5$ GeV$^{-1}$, which makes the experimental
search for $Y(1915)$ become possible, where its main decay modes
include $\rho a_0(980)$, $b_1(1235)\pi$, $\rho(1450)\pi$,
$\rho\pi$, $f_0(980)\omega$ and $KK$ (see the left diagram of Fig.
\ref{Fig: width1} for more details). Just because $Y(1915)$ is
predicted as the non-strange partner of $Y(2175)$, a
realistic experimental study of $Y(1915)$ can be via the
$e^+e^-\to f_0(980)\omega$ or $e^+e^-\to f_0(980)\pi\pi$ process,
which is supported by our calculation, where the decay width of
$Y(1915)\to f_0(980)\omega$ is around $5.86\sim16.22$ MeV. Besides
directly producing $Y(1915)$ by the $e^+e^-$ collision, the BESIII
experiment can be as an ideal platform to search for $Y(1915)$,
{\it i.e.}, the study of the $f_0(980)\omega$ invariant mass
spectrum of the $J/\psi\to \eta f_0(980)\omega$ decay.
Additionally, we also find several partial decay width ratios
\begin{eqnarray*}
&&\frac{\Gamma(b_1(1235)\pi)}{\Gamma(\rho(1450)\pi)}\approx1.3\sim3.5,\; \frac{\Gamma(b_1(1235)\pi)}{\Gamma(a_0(980)\rho)}\approx0.6\sim 1.7,\\
&&\qquad\qquad \quad\frac{\Gamma(a_0(980)\rho)}{\Gamma(f_0(980)\omega)}\approx9\sim 19,
\end{eqnarray*}
which are not too dependent on the variation of $R$ value. To some extent, these ratios can be served as the further experimental test to the predicted $Y(1915)$.

For $\omega(1650)$, the calculated total decay width when taking $R=3.5\sim 4.5$ GeV$^{-1}$ is overlap with the result given by Achasov {\it et al.} in Ref. \cite{Achasov:2003ir} if considering the experimental error. Checking the data of $\omega(1650)$ listed in Particle Data Group (PDG) indicates that different experiments gave the
experimental widths different from with each other. Our calculation partly supports the measurement result of $\omega(1650)$ in Ref. \cite{Achasov:2003ir}, where $\omega(1650)$ is a state with broad width. The $\omega(1650)$ results presented in Fig. \ref{Fig: width1} also show that $b_1(1235)\pi$ is its dominant decay channel, which almost contributes to the total width of $\omega(1650)$. In addition, its main decay modes include $\rho\pi$, $\omega\eta$, $KK$ and $K^*K$, where $\omega(1650)$ decays into $\rho\pi$ and $\omega\eta$
were seen in experiments \cite{Nakamura:2010zzi}.
Thus, the results in Fig. \ref{Fig: width1} also support $\omega(1650)$ as a $1D$ $\omega$ state.
Furthermore, our study of $\omega(1650)$ raises one issue: further experimental measurement of the resonance parameter for $\omega(1650)$ will be helpful to clarify the present mess of total width of $\omega(1650)$, where we are inclined to $\omega(1650)$ as a broad state according to our investigation, which can be tested by future experiment. Here, we also listed some partial decay width ratios, which do not strongly depend on the model parameters, {\it i.e.}, $\Gamma(b_1(1235)\pi)/\Gamma(\rho\pi)\approx5.9\sim8.8$, $\Gamma(\rho\pi)/\Gamma(\omega\eta)\approx2.9\sim3.9$,
$\Gamma(\rho\pi)/\Gamma(KK^*)\approx7.1\sim9.3$, $\Gamma(b_1(1235)\pi)/\Gamma(KK^*)\approx55.0\sim62.3$, $\Gamma(\omega\eta)/\Gamma(KK^*)\approx2.3\sim2.4$, $\Gamma(KK^*)/\Gamma(KK)\approx0.7\sim1.2$ and $\Gamma(\rho\pi)/\Gamma(KK)\approx6.4\sim8.5$.

There exist five decay channels open for $\omega(1420)$ listed in Fig. \ref{Fig: width1}. Among these decay channels, $\rho\pi$, $b_1(1235)\pi$, $\omega\eta$ and $KK$ are the main decay channels, though these decay widths are dependent on parameter $R$. As indicated in PDG, $\rho\pi$ channel is the dominant decay channel for $\omega(1420)$, which is confirmed by our calculation. Besides, $b_1(1235)\pi$ was seen in experiment, which is also supported by our result. When comparing the total width with the experimental data [see Fig. \ref{Fig: width1}], we find that our result is comparable with the BaBar recent data, where the measured width of $\omega(1420)$ is $130\pm50\pm100$ MeV by analyzing the $e^+e^-\to \omega \pi^-\pi^-\gamma$ process \cite{Aubert:2007ef}. Thus, $\omega(1420)$ as $2S$ state is further confirmed by our phenomenological study.

\begin{center}
\begin{figure}[htbp!]
\includegraphics[bb=5 160 700 680,scale=0.55,clip]{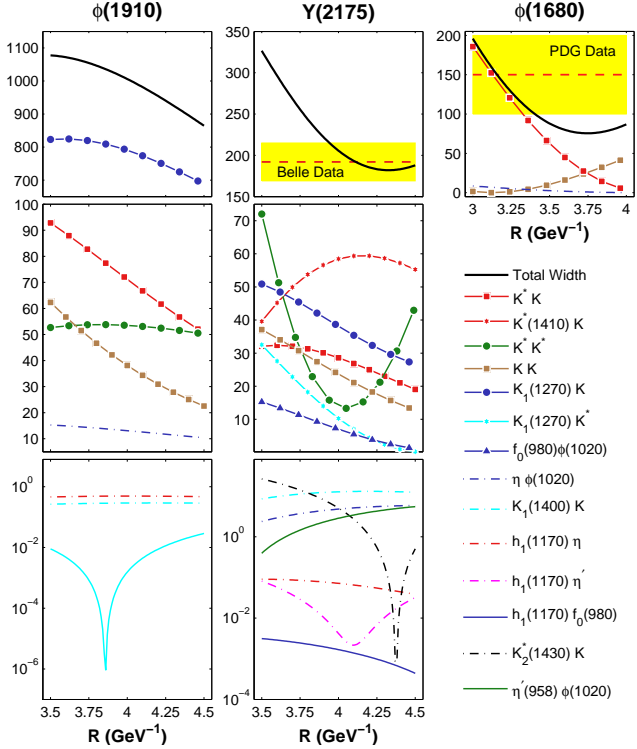}
\caption{(color online). The total and partial decay widths of
$\phi(1910)$, $Y(2175)$ and $\phi(1680)$ as strangeonia with $1D$,
$2D$ and $2S$ quantum numbers, respectively. \label{Fig: width2}}
\end{figure}
\end{center}

In the following, we illustrate the decay behavior of
$\phi(1910)$. At present, vector strangeonium with $1D$ is still
absent in experiment. If the predicted $\phi(1910)$ is the
candidate for such a $1D$ state, our calculation indicates that
$\phi(1910)$ is a very broad state with total width around
$822\sim 1047$ MeV just shown in Fig. \ref{Fig: width2}. Because
of the difficulty of finding very broad structure in experiment,
it naturally explains why the present experiment has not observed
any evidence of vector strangeonium with $1D$ quantum number. If
one experimentally searches $\phi(1910)$, the result presented in
Fig. \ref{Fig: width2} shows that $\phi(1910)$ mainly decays into
$K_1(1270)K$, $K^*K$, $K^*K^*$, $KK$ and $\eta \phi(1020)$, where
$K_1(1270)K$ channel is the dominant decay of $\phi(1910)$ since
the branching ratio of $\phi(1910)\to K_1(1270)K$ can reach up to
$(79\sim84)\%$. Apart from the total width information of
$\phi(1910)$, we also find some $\phi(1910)$ partial decay width
ratios, {\it i.e.},
$\Gamma(K_1(1270)K)/\Gamma_{total}\approx0.66\sim0.85$,
$\Gamma(KK^*)/\Gamma(KK)\approx0.8\sim2.6$,
$\Gamma(K^*K^*)/\Gamma(\eta\phi)\approx2.54\sim5.71$ and
$\Gamma(K_1(1270)K)/\Gamma(KK^*)\approx5.32\sim16.27$, which are
stable in the range of $R$ discussed here. These information is
also valuable for further experimental investigation of
$\phi(1910)$.

As a test of the quark pair creation model, one can see that the result of
$\phi(1680)$ is consistent with the existing experimental
measurement. Calculation of $\phi(1680)\to K^*K$ confirms $K^*K$
as the dominant decay channel of $\phi(1680)$ given by experiment
\cite{Nakamura:2010zzi}. In addition, when taking $R=3.4$
GeV$^{-1}$, the calculated $\Gamma(KK)/\Gamma(K^*K)=0.07$ is in
agreement with experimental value listed in PDG while the
corresponding total width are $93$ MeV, which is close to the
lower limit of the PDG data (see Fig. \ref{Fig: width2} in more
details). In reality, there should exist some small difference between the $R$ values for $\phi$ and $\omega$. In our calculation shown in Figs. \ref{Fig: width1} and \ref{Fig: width2}, however, instead of taking the definite values of $R$ to give the corresponding results, we have given the trend of these widths with a certain range of $R$. Here we take the same range of $R$ when discussing $\phi$ and $\omega$ with the same quantum number.

For clarifying whether the observed $Y(2175)$ can be explained as the conventional strangeonium, we also calculate the total and partial widths of $Y(2175)$ with the assignment of $2D$ vector strangeonium. We can find the overlap of theoretical and Belle results in the range of $R=3.94\sim4.84 $ GeV$^{-1}$, which is consistent with the estimate of $R$ value for $2D$ $s\bar{s}$ states \cite{Close:2005se}. However, the calculated total width is far larger than the width measured by BaBar \cite{Aubert:2006bu} and BES \cite{:2007yt}. If adopting the Belle measurement, we can conclude that it is reasonable to explain $Y(2175)$ as a $2D$ state in $\phi$ family. Of course, the measurement of its resonance parameter will be helpful to further test this assignment to $Y(2175)$. In Fig. \ref{Fig: width2}, we also give the main decay modes of $Y(2175)$, which can provide important information to experimental search for $Y(2175)$ through its other decays, where ratios $\Gamma(K_1(1270)K)/\Gamma(KK)\approx1.5\sim2.3$ and $\Gamma(K^*K^*)/\Gamma_{total}\approx0.1\sim0.6$ weakly depend on the model parameter.

\begin{center}
\begin{figure}[htbp!]
\includegraphics[bb=5 160 700 680,scale=0.53,clip]{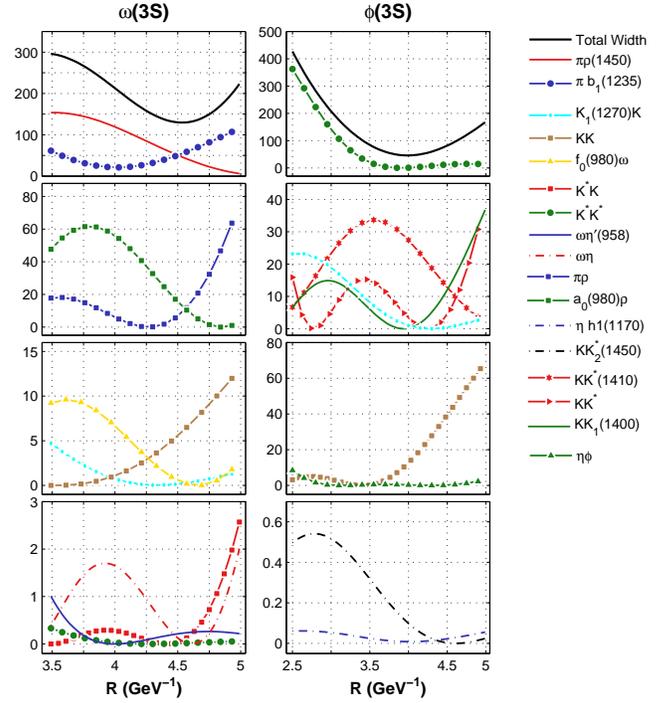}
\caption{(Color online.) The total and partial decay widths of
$\omega(3S)$ and $\phi(3S)$. \label{Fig: width3}}
\end{figure}
\end{center}

By the Regge trajectories analysis, we can predict the masses of $\omega$ and $\phi$ with $3S$ quantum numbers, {\it i.e.,} $m_{\omega(3S)}=1830$ MeV and $m_{\phi(3S)}=1940$ MeV. In Fig. \ref{Fig: width3}, their strong decay behaviors are given. For the $\omega(3S)$ meson, the total width is about $130\sim 296$ MeV corresponding to $R=3.5\sim 5$ GeV$^{-1}$. The dominant decays are $\pi\rho(1450)$, $\pi b_1(1235)$, and $a_0(980)\rho$. For $\phi(3S)$, the total decay width is strongly dependent on the value $R$ due to the node effects of the wave function, which is reflected by Fig. \ref{Fig: width3}. Among the all two-body strong decay channels, $K^*K^*$ is the dominant decay channel. In addition, $KK^*(1410)$ is the main decay chanel of $\phi(3S)$.

In summary, the similarity between the $\phi$ and $\omega$ families enables the study of these meson families to be borrowed from each other. Stimulated by the observation of strangeonium-like state $Y(2175)$, we first predict its non-strange partner $Y(1915)$ by the mass spectrum analysis, where $Y(1915)$ with other states listed in Eq. (\ref{common}) seems to form a complete series of flavors. The study of $Y(1915)$ decay behavior indicates that $Y(1915)$ is a broad state with width around $317\sim 354$ MeV. Considering its main decay modes, we further propose that $e^+e^-\to f_0(980)\omega, f_0(980)\pi\pi$,  or $J/\psi\to \eta f_0(980)\omega$ can be as the realistic process of searching for the predicted $Y(1915)$. Besides the prediction of $Y(1915)$, we also obtain the information on $1D$ state in the $\phi$ family by
the comparison of mass spectra of $\phi$ and $\omega$ families. Our study shows that this $1D$ state
is a very broad resonance structure with mass of about 1910 MeV, which naturally explains why the present experiment have not found any evidence of this state since it is not easy to identify broad structure in experiment. As the strange partner of the predicted $Y(1915)$, $Y(2175)$ could be explained as a $1D$ strangeonium. The phenomenological study presented in this letter is not only helpful to reveal the underlying properties of these light hadrons, but also will serve further experimental investigation.

\vfill

\noindent{\bf Acknowledgment} X.L. would like to thank Dr.
Cheng-Ping Shen for useful discussion of the Belle measurement.
This project is supported by the National Natural Science
Foundation of China under Grant Nos. 11175073, 11005129, 11035006,
the Ministry of Education of China (FANEDD under Grant No. 200924,
DPFIHE under Grant No. 20090211120029, NCET, the Fundamental
Research Funds for the Central Universities), the Fok Ying Tung Education Foundation (No. 131006), and the West Doctoral Project of Chinese Academy of Sciences.

\end{document}